\documentclass[traditabstract,a4,10pt]{aa}

\usepackage{graphicx}
\usepackage{natbib,amsmath,amssymb}

\usepackage{longtable}

\newcommand{\elie}{{\mathbf E}}
\newcommand{\elih}{{\mathbf H}}

\newcommand{\ka}{\kappa^{q_1q_2}}
\newcommand{\kax}{\kappa^R}
\newcommand{\kc}{\kappa^{xy}}
\newcommand{\kps}{\kappa^{R\theta}}
\newcommand{\krphi}{\kappa^{r\phi}}
\newcommand{\krz}{\kappa^{Rz}}

\newcommand{\domain}{{\Omega'}}
\newcommand{\pdomain}{{\cal S}}
\newcommand{\csection}{{\cal S}}

\newcommand{\grav}{{\cal G}}

\newcommand{\hpsi}{{\cal H}}
\newcommand{\hhpsi}{{\cal U}}
\newcommand{\spsi}{{\cal H}}

\newcommand{\asinh}{\,{\rm asinh \;} }
\newcommand{\atanh}{\,{\rm atanh \;} }
\newcommand{\atan}{\,{\rm atan \;} }
\newcommand{\qu}{{q_1}}
\newcommand{\qd}{{q_2}}
\newcommand{\fdeux}{{f_2}}
\newcommand{\gdeux}{{g_2}}
\newcommand{\qt}{{q_3}}
\newcommand{\rp}{{r'}}
\newcommand{\dep}{{\Delta'}}
\newcommand{\des}{{\delta}}
\newcommand{\slfrac}[2]{\left.#1\middle/#2\right.}
\newcommand{\dd}{\delta}
\newcommand{\cotan}{\mathrm{cotan} \;}

\newcommand{\elik}{{\mathbf K}}
\newcommand{\elipi}{{\mathbf \Pi}}        
\newcommand{\pdz}{\partial_Z}

\begin{document}

\title{Exact, singularity-free recasting\\ of the Newtonian potential in continuous media}
\authorrunning{J.-M. Hur\'e}

\titlerunning{Exact, singularity-free recasting of the Newtonian potential}

\author{Jean-Marc Hur\'e\inst{1,2}}

\offprints{jean-marc.hure@obs.u-bordeaux1.fr}

\institute{Universit\'e de Bordeaux, OASU, 351 cours de la Lib\'eration, F 33405 Talence
\and
CNRS, UMR 5804, LAB, 2 rue de l'Observatoire, BP 89, F 33271 Floirac\\
\email{jean-marc.hure@obs.u-bordeaux1.fr}}

\date{Received ??? / Accepted ???}

\abstract{
The gravitational potential is a key function involved in many astrophysical problems. Its evaluation inside continuous media from Newton's law is known to be challenging because of the diverging kernel $1/|\vec{r}-\vec{r}'|$. This difficulty is generally treated with avoidance techniques (e.g. multipole expansions, softening length) themselves not without drawbacks. In this article, we present a new path that basically fixes the point-mass singularity problem in systems with, at least, two dimensions. It consists of recasting the gravitational potential $\psi$ in an equivalent integro-differential form, 
$$\psi(\vec{r}) = \frac{1}{f(\vec{r})} \partial^2_{\qu \qd} \spsi(\vec{r}),$$
where $(\qu,\qd)$ is a pair of independent spatial variables (linear and/or angular), $f$ is a known function, and $\spsi$ is an auxiliary scalar function. In contrast with $\psi$, this ``hyperpotential'' $\spsi$ is the convolution of the mass density with a {\it finite amplitude} kernel $\kappa$. We show that closed-form expressions for $\kappa$ can be directly deduced from the potential of homogeneous sheets. We then give a few formulae appropriate to the Cartesian, cylindrical and spherical coordinate systems, including axial symmetry. The method is essentially not limited, either on the geometry of the source or on the distribution, and  its implementation is straightforward. Several tests based upon simple quadrature/differentiation schemes are presented (the homogeneous rectangular sheet, cuboid and disk, the Maclaurin disk and a truncated Lane-Emden solution). Compared with a direct summation, the extra computational cost is low and the gain is real: no truncated series, no free parameter, and a relative accuracy better than $1 \%$ for typically $16$ nodes per spatial direction using the most basic numerical schemes. 
}


\maketitle

\section{Introduction}

Gravitation plays an essential role in the evolution of most astrophysical systems, from aggregates and dusty planetary rings to rotating stars, supermassive black holes in active nuclei and galactic clusters \citep{hachisu86,ko04,col06,king10,co11}. In the investigation of various dynamical problems and equilibrium configurations from first integrals and energy equations, the potential appears as a fundamental scalar function. In continuous media, it is naturally accessible through an integral, namely
\begin{equation}
\psi(\vec{r})=-\grav\int{\frac{dm'}{|\vec{r}-\vec{r}'|}},
\label{eq:psi}
\end{equation}
where the kernel sweeps aways the point mass singularity | a direct consequence of Newton's inverse square law. Indeed, the potential integral is convergent for most density distributions of physical interest \citep{kellogg29,durand64,binneytremaine87}. The singularity problem is inherent in the discretization-counting technique usually adopted. By dividing the system into small massive elements and summing over all individual contributions, it is difficult to estimate precisely the influence of any small element upon itself (i.e., ``self-gravity''), which is not ameliorate a lot by increasing the resolution.

The point mass singularity can be avoided in various ways that are more or less faithful to Newton's law. The multipole expansion of the kernel $|\vec{r}-\vec{r}'|^{-1}$ one of the most valuable theoretical tools in potential theory \citep{kellogg29,durand64,cohl01,ak99}. It is extremely efficient outside the material domain and a few terms often suffice to reach computer precision. Inside and even in close neighborhood, however, the convergence of the series is known to be poor because $r/r' \approx 1$. Because the series is an alternate series, convergence is much delayed and truncations are critical \citep{clement74}. Low convergence is a common property of multipole expansions and is observed in various contexts other than gravitation \citep{wu75,ko20,gra10}. Users of multipole expansions generally need to incorporate a large number of terms | tens to hundred typically | before accuracy becomes acceptable \citep{hachisu86,stonenorman92,machm12}. Because the number of integrals to estimate is equal to the number of terms, the computational time increases linearly. Another option to derive $\psi$ is the Poisson equation, which is rapidly solved with specific algorithms \citep{stonenorman92,sto93,Spotz95high-ordercompact,Briggs00,maha03,jusu07,gt11}. Nevertheless, Poisson-solvers are not always ``self-starting'', meaning they require precise boundary conditions {\it only the integral approach can furnish}. Another drawback is the shape of astrophysical bodies, which are often complex and not systematically match the numerical meshes \citep[for techniques based on mapping, see ][]{gc01,reese06}. In particular, the Poisson equation is three-dimensional by nature, and not well-suited to problems in one and two dimensions.

In this paper, we describe a novel path for determining the Newtonian potential of a continuous system by recasting Eq.(\ref{eq:psi}). The new form does not involve any singular kernel, series, or ``softening length'', but just the cross-derivative of the mass density convolved with a {\it finite amplitude kernel}. The recasting is exact and general in the sense that i) it preserves the Newtonian character of the interactions at all scales, and ii) it applies to any density distribution and morphology (shape and number of dimensions larger than one). This is therefore a new tool for both numerical applications and theoretical investigations in various domains of astrophyscis (e.g., simulations, generation of approximations, determination of potential/density pairs) and Physics as well. This paper goes beyond the analysis presented in \cite{hd12}, which was restricted to axial symmetry. We consider here i) a generic treatment of the regularization step, regardless of the system of coordinates, ii) a full three-dimensional approach, iii) a simple recipe to determine the finite amplitude kernel, and iv) a direct application to the Cartesian, cylindrical, and spherical coordinates. 

The paper is organized as follows. In Section \ref{sec:pi}, we recall the integral expression for the Newtonian potential of a continuous distribution. We formally describe the recasting of the potential integral based upon the properties of Newton's law (symmetry and independant spaces). The application to the Cartesian, cylindrical, and Spherical coordinate systems is the aim of Section \ref{sec:res}. In particular, we illustrate the method by considering a few test-cases, mostly of astrophysical interest, by using deliberately low-order numerical schemes (our goal is not to perform a critical study of the most efficient techniques for quadratures and differentiations). A conclusion summarizes the results and mentions possible issues to consider next. A few appendices contain formulae and demonstrations.
 
\section{Recasting of the potential integral}
\label{sec:pi}

\begin{figure}
\centering
\includegraphics[width=8.cm]{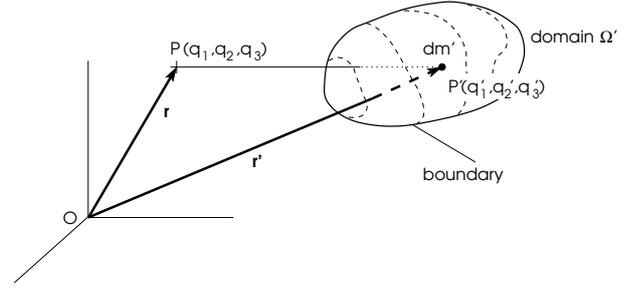}
\caption{Field point P$(\vec{r})$, source point P$'(\vec{r}')$ belonging to the material domain $\domain$, and elementary mass $dm'$. Inside $\domain$, the separation PP$'$ vanish.}
\label{fig:vol}
\end{figure}

The Newtonian potential at a point P$(\vec{r})$ in space of a body is given by Eq.(\ref{eq:psi}). The integral extends over the material domain  $\domain$ (including its boundary), i.e., $\psi(\vec{r}) \equiv \psi(\vec{r};\domain)$, $dm'(\vec{r}')$ is the elementary mass at P$'(\vec{r}') \in \domain$, $|\vec{r}-\vec{r}'|=$PP$'$, and $\grav$ is the constant of gravity. The configuration is illustrated by Fig. \ref{fig:vol}. As is well known, PP$'$ vanishes everywhere inside  $\domain$, making the kernel singular, while the potential is, most of the time, a finite function of space \citep[see e.g.][]{kellogg29}.

\subsection{Idea behind recasting and strategy}
\label{subsec:idea}

As Eq.(\ref{eq:psi}) shows, there are two spaces in potential theory: i) the space of {\it field points} where the potential is requested (hereafter, the P-space), and ii) the space of {\it source points} that describes the source (hereafter, the P$'$-space). These spaces are superimposed in practice | this is the physical space |, but are {\it decoupled} mathematically. Indeed, when estimating the potential from Eq.(\ref{eq:psi}), $\vec{r}$ is held fixed while the integration is performed in the P$'$-space. The point-mass singularity, of hyperbolic-type, can be regularized using {\it two successive integrations in orthogonal directions} (as a proof, note that the potential of flat or curved homogeneous sheets is generally a finite function of space and source parameters). The idea is then to integrate the Newton kernel in the P-space until the singularity is finally suppressed. This operation is necessarily possible since the {\it Newton kernel is symmetrical} with respect to $\vec{r}$ and $\vec{r}'$. Concretely, if P has coordinates $(\qu,\qd,\qt)$, then the {\it regularization step} becomes\footnote{Under invariance, the hyperbolic singularity can be converted into a logarithmic singularity that is subsequently regularized by a single integration, i.e., $\int{\frac{fd\qu}{|\vec{r}-\vec{r}'|}}$, but this is a special case (see Sect. \ref{sec:res}).}
\begin{equation}
\iint_{\qu,\qd}{\frac{f(\vec{r}) d\qu d \qd}{{\rm PP}'(\qu,\qd,\qt)}} \equiv \ka(\vec{r};\vec{r}'),
\label{eq:k0}
\end{equation}
where $f$ is introduced for convenience (see below) and it is a function of P only. At this stage, the coordinates $(\qu',\qd',\qt')$ of P$'$ are regarded as parameters, and $\ka$ must be a function of $\vec{r}$. Since the regularization is performed in the P-space, it is made {\it regardless of the mass distribution}, which is especially attractive. The new kernel $\ka$ (hereafter the ``hyperkernel'') has, by construction, a {\it finite amplitude} and can be convolved with the mass density. This is the {\it convolution step}:
\begin{equation}
- \grav \int_\domain{\ka(\vec{r};\vec{r}') dm'} \equiv \spsi(\vec{r};\domain),
\label{eq:hyperkernel}
\end{equation}
where the factor $- \grav$ is introduced for convenience (see below). This integral produces an auxiliary scalar function, $\spsi$ (hereafter,  the ``hyperpotential''). The Newtonian potential is then recovered by reversing the regularization-step. This is the {\it recovering step}:
\begin{flalign}
\label{eq:pah}
\partial^2_{\qu\qd} \spsi & = - \grav \partial^2_{\qu\qd}{\int_\domain{\ka dm'}},\\
\nonumber
 & = - \grav \int_\domain{\left( \partial^2_{\qu\qd} \ka \right) dm'},\\
\nonumber
&= - \grav f \int_\domain{ \frac{dm'}{|\vec{r}-\vec{r}'|} }\\
\nonumber
&= f \psi.
\end{flalign}
The advantage of this approach is twofold: the singularity is circumvented, and at the same time, it is accounted for exactly. In practice, the absence of diverging kernel renders step 2 easier than with the Newton kernel. Because convolutions produce smooth functions, step 3 is also expected to be uncomplicated. Step 1 is by far the most critical, but {\it it is made once only} provided the hyperkernel is {\it analytical} (there is no interest in the recasting if $\ka$ is to be determined by numerical means). The gravitational potential is finally found from steps 2 and 3. The extra-cost is therefore low: there is only an additional differentiation compared to the classical approach, but the singularity is correctly managed.
 
\subsection{Note on Chandrasekhar and Lebovitz superpotentials}

Our approach may evoke some aspects of the theory developped in \cite{cl62} and subsequent papers \citep[see also][]{chandra87}. These authors have shown that the Newtonian potential is the trace of a symmetric tensor potential, but is also the source term of a Poisson equation, namely
\begin{equation}
\nabla^2 \chi = -2 \psi,
\end{equation}
where the solution is
\begin{equation}
\chi \equiv - \grav \int_\domain{|\vec{r}-\vec{r}'|  dm'}.
\end{equation}
It is clear that $\spsi$ and $\chi$ (called ``superpotential'') share two common properties: i) they are the convolution of the density field by a {\it finite amplitude kernel}, and ii) they exactly reproduce the gravitational potential by partial differencing. However, the present recasting differs from the theory of superpotentials on the following points:
\begin{itemize}
\item[i)] $\psi$ is determined here through a single second-order partial derivative (not three);
\item[ii)] the recasting is not limited to $3$D-problems, but works for $2$D-problems as well;
\item[iii)] it is not really specific to a particular system of coordinates, while Cartesian coordinates are mostly used in \cite{cl62}.
\end{itemize}

In some sense, the hyperpotential $\spsi$ is some kind of optimized version of the $\chi$-function, especially designed for numerical applications which is, initially, our main motivation.

\begin{figure}
\centering
\includegraphics[width=8.8cm]{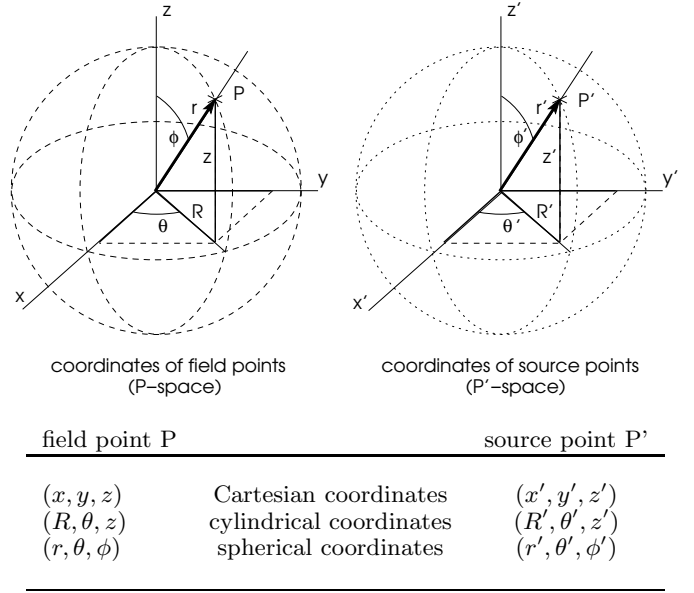}
\begin{tabular}{lcc}\\
field point P & &  source point P' \\ \hline \\
$(x,y,z)$ & Cartesian coordinates &  $(x',y',z')$ \\
$(R,\theta,z)$ & cylindrical coordinates &  $(R',\theta',z')$\\
$(r,\theta,\phi)$ & spherical coordinates &  $(r',\theta',\phi')$ \\ \\\hline
\end{tabular}
\caption{Notations for the Cartesian, cylindrical, and spherical coordinates: P-space ({\it left}) and P$'$-space ({\it right}).}
\label{fig:ccs}
\end{figure}

\begin{table*}
\centering
\begin{tabular}{lccccc}
                          &  $\qu, \qd$    &$f$ & $d^2A$  & surface $\pdomain$ & hyperkernel $\ka$ \\ \hline  \hline\\
Cartesian coordinates     &   $x,y$         &$1$ & $dxdy$         & rectangular sheet & $\kc$, see Eq.(\ref{eq:k0xy}) \\
                          &   $y,z$         &$1$ & $dydz$         & rectangular sheet & $\kappa^{yz}$ (see $\kc$)\\
                          &   $x,z$         &$1$ & $dxdz$         & rectangular sheet & $\kappa^{xz}$ (see $\kc$)\\\\
Cylindrical coordinates   &   $\theta,z$    &$R$ & $Rd\theta dz$  & piece of hollow cylinder & | \\
                          &   $R,\theta$    &$R$ & $R d\theta dR$ & polar sector & $\kps$, see Eq.(\ref{eq:k0atheta})\\
                          &   $R,z$         &$1$ & $dR dz$        & meridional sheet  & $\krz$, see Eq.(\ref{eq:ms})\\ \\
Spherical coordinates     &   $\theta,\phi$ & $r^2 \sin \phi$ & $r^2 \sin \phi d\phi d\theta$ & spherical cap & |\\
                          &   $r, \phi$     &$r$ & $r dr d\phi $  & meridional sector & $\krphi$, see the Appendix \ref{app:msect} \\
                          &   $r,\theta$    &$r \sin \phi$ & $r \sin \phi dr d\theta$ & piece of cone & | \\ \\ \hline \\
Axial symmetry$^*$            &   $R$           &$R$ & $dR$     & disk &  $\kax$, see Eq.(\ref{eq:k0_as})\\
                          &   $r$           &$1$ & $dr$     & cone & | \\
                          &   $\phi$        &$1$ & $d\phi$   & spherical cap & | \\  \\ \hline \hline
\end{tabular}
\caption{Pairs $(\qu,\qd)$, function $f$, and associated area element $d^2A$ for the three most popular coordinate systems. The formula for the hyperkernel $\kappa^{\qu\qd}$, when known in closed form, is indicated in the last column (otherwise '|'). $^*$Under axial symmetry, a single variable is necessary (see Sect. \ref{subsec:asym}).}
\label{tab:kappas}
\end{table*}

\section{Results and examples}
\label{sec:res}

\subsection{Derivation of hyperkernels, and the link with the potential of homogeneous sheets}
\label{subsec:tne}

According to Eq.(\ref{eq:k0}), the recasting depends on the capability to determine analytically an expression for $\ka$ associated with a given pair $(\qu,\qd)$ of coordinates, preferably a closed-form. There are many possibilities, in particular because of the presence of the function $f(\qu,\qd,\qt)$, which adds a degree of freedom. If we define $f$ such that
\begin{equation}
d^2A = f(\qu,\qd,\qt) d\qu d\qd
\label{eq:ae}
\end{equation}
is an {\it area element}, then
\begin{equation}
\ka = \iint_{\pdomain}{\frac{d^2A({\rm P})}{|\vec{r}-\vec{r}'|}},
\label{eq:k0fromda}
\end{equation}
where $\pdomain$ is a surface $\qt=const$ in the P-space. We see that Eq.(\ref{eq:k0fromda}) is nothing but, up to a factor $-\grav$, the formula for the {\it gravitational potential of a homogeneous surface with unit surface density $dm/d^2A$}, except that the role of the P-space and P$'$-space is exchanged. To obtain a formula for $\ka$, it is sufficient to extract from the list of known potential/density pairs those that correspond to a bi-dimensional distribution (i.e., {\it a sheet}) and constant surface density. As Eq.(\ref{eq:k0fromda}) suggests, there is no special constraint on the shape and size of the sheet (i.e., flat, curved, rectangular, circular, etc.). Nevertheless, from a practical point of view, it seems preferable that $\pdomain$ and $\domain$ be geometrically ``compatible'' enough to facilitate the convolution-step. Since there is certain freedom in selecting the integral bounds in Eq.(\ref{eq:k0fromda}), it is also better to consider a finite size (and finite mass) sheet.

\subsection{An easy implementation}

For each point P where $\psi$ is requested, the sequence of operations is the following:
\begin{enumerate}
\item computing the hyperkernel $\ka$ from an appropriate formula (this depends on the coordinate system; see next section);
\item estimating the hyperpotential $\spsi$ from Eqs.(\ref{eq:hyperkernel}) for the actual density distribution (a volume density $\rho$ or a surface density $\Sigma$). A quadrature scheme is needed. This is the first place where numerical errors are generated;
\item estimating the cross-derivative of $\spsi$. A differentiation scheme is needed. This requires determining hyperpotential-values in the vicinity of the actual point P. This is the second place where numerical errors are generated. 
\end{enumerate}

Clearly, all techniques curently used to compute $\psi$ from Eq.(\ref{eq:psi}) can be employed for $\spsi$. Because $\spsi$ is a convolution (see below), fast specific algorithms coupled with high-perfomance differentiation schemes can probably be envisaged. Tests presented in the following will employ the most basic schemes, which have produced good results.

\begin{figure}
\centering
\includegraphics[width=8.7cm,bb=31 8 770 583,clip==]{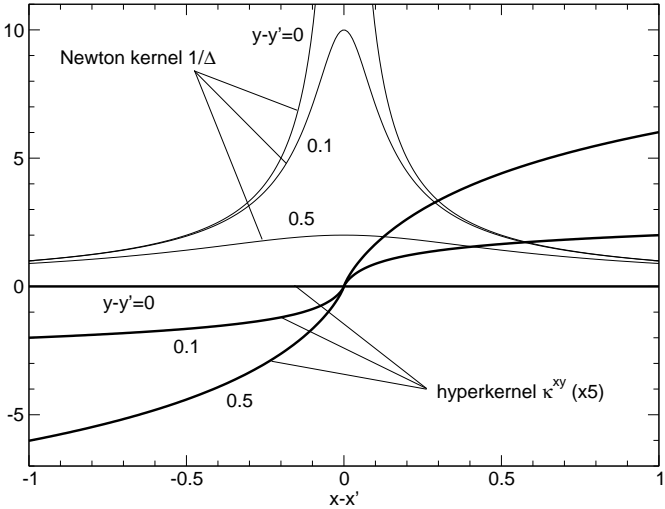}\\
\caption{The hyperkernel $\kc$ in the $(x,y)$-plane in the vicinity of point P$'$ (magnified by a factor $5$), for $z=z'$ and three different values of $y-y'$ labelled on the curves. The Newton kernel which diverges for $|\vec{r}-\vec{r}'|=0$ is shown for comparison.}
\label{fig:cart}
\end{figure}

\begin{figure}
\centering
\includegraphics[width=8.7cm,bb=70 50 320 264,clip==]{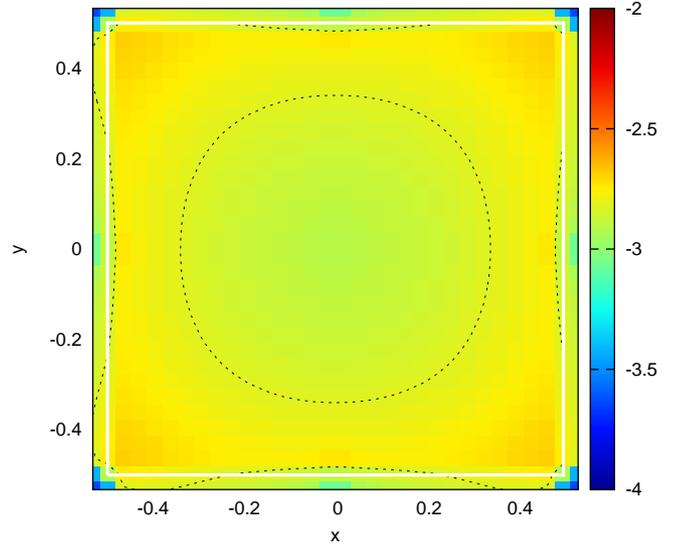}
\caption{Error index $\epsilon$ on potential values in the plane of the homogeneous square sheet with vertices at $(\pm \frac{1}{2},\pm \frac{1}{2},0)$ (boundary in white). The mean error index is $-2.52$ (dashed line).}
\label{fig:epsi_cart_sheet.ps}
\end{figure}

\subsection{Results. The case of Cartesian coordinates}
\label{subsec:cart}

Table \ref{tab:kappas} lists pairs $(\qu,\qd)$, the associated area element $d^2A$, and the function $f$ appropriate for the Cartesian, cylindrical and spherical coordinates that are most often used. Other coordinate systems and geometries can obviously be considered as well. The notations are summarized in Fig. \ref{fig:ccs}. In Cartesian coordinates, the recasting that corresponds to the pair $(\qu,\qd)=(x,y)$ is
\begin{equation}
\psi(\vec{r}) = \partial^2_{xy}{ \int_\domain{\kc dm'}},
\label{eq:psifinalxy}
\end{equation}
and other pairs can be considered by permutation. As argued above, we can determine $\kc$ by considering a surface $\qt \equiv z=const$, i.e., a flat horizontal sheet, and the most natural choice is the rectangular shape. The formula for the potential is known in that case \citep[e.g.][]{durand64}. It is reproduced in Appendix \ref{app:rectsheet}. The hyperkernel is deduced by exchanging P and P$'$, which leads to Eq.(\ref{eq:k0xy}). Figure \ref{fig:cart} displays $\kc$ in the $(x,y)$-plane in the vicinity of the point P$'(x',y')$, for three different values of $y-y'$ and for $z-z'=0$. The Newton kernel $1/|\vec{r}-\vec{r}'|$ is also shown for comparison. We can see that the hyperkernel is a smooth function with finite amplitude. In particular, it is zero at zero relative separation. Any type of mass density profile can be injected in Eq.(\ref{eq:psifinalxy}), bi- or three-dimensional, uniform or not.

We illustrate the potential recasting with two simple examples. As a first test, we consider a square sheet in the $(x,y)$-plane with constant surface density $\Sigma_0$ (or $dm'=\Sigma_0 dx'dy'$), length unity, and centered on the origin, with vertices at $(\pm \frac{1}{2},\pm \frac{1}{2},0)$. It is discretized on a $N' \times M'$ grid with regular spacing in each direction. The P-grid is made of $N \times M$ points with uniform spacing as well. The quadratures and partial derivatives are determined by second-order schemes that are among the most basic ones \citep[e.g.][]{numrec92}. Figure \ref{fig:epsi_cart_sheet.ps} shows the error index
\begin{equation}
\epsilon = \log \left[ \max \left(2 \times 10^{-16}, \left| 1 - \frac{\psi}{\psi_e}\right| \right) \right]
\label{eq:epsilon}
\end{equation}
between $\psi$ determined from Eq.(\ref{eq:psifinalxy}) and the reference potential $\psi_e$ ($2 \times 10^{-16}$ is for double-precision computations). This case is for $N'=M'=32$ and $N=M=N'+2$, which leaves just one point outside the sheet, left and right, bottom and top. We see that the relative error is rather uniform inside the material domain, of about $0.1 \%$. Close to the edges of the sheet, the error rises slightly. Outside $\domain$, the accuracy is still uniform, but typically better by an order of magnitude. 

\begin{figure}
\centering
\includegraphics[width=8.7cm,bb=70 50 320 264,clip==]{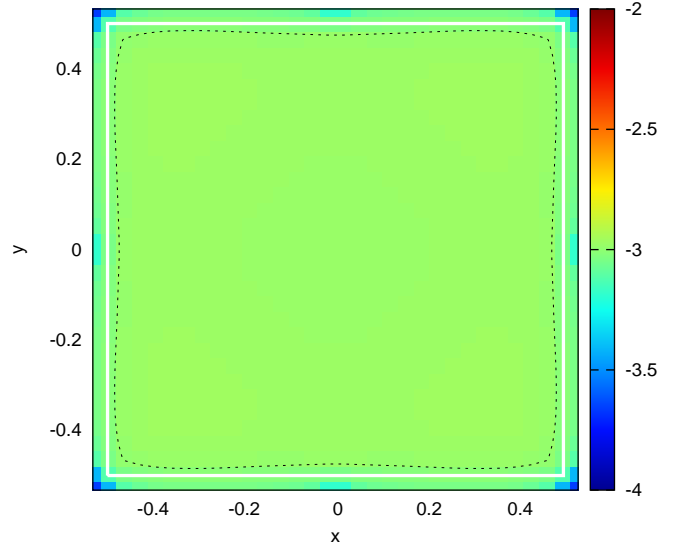}
\caption{Same conditions and same color code as for Fig. \ref{fig:epsi_cart_sheet.ps}, but for the homogeneous cuboid with vertices at $(\pm \frac{1}{2},\pm \frac{1}{2},\pm \frac{1}{2})$. The mean value is $-3.00$ (dashed line).}
\label{fig:epsi_cart_cuboid}
\end{figure}

As a second test, we consider a cube with uniform density $\rho_0$ (i.e., $dm'=\rho_0 dx'dy'dz$), length unity, and vertices at $(\pm \frac{1}{2},\pm \frac{1}{2},\pm \frac{1}{2})$. The numerical setup is the same as for the sheet, and $z=0$. The reference potential is also known for this $3$D body \citep{macmillan1930theory,wal76}. Figure \ref{fig:epsi_cart_cuboid} displays the error index versus $x$ and $y$ in the cube's midplane. Again, we notice that the relative error is uniform, with $0.2 \%$ typically inside the body, and a factor $10$ better outside. Edge effects are less marked than in $2$D. The integration of the kernel in the third direction smoothes the errors, and the potential is now derivable when crossing the lateral faces of the cube. The Fortran 90 program used in these two examples is available upon request.

\subsection{The hyperpotential is a convolution product}

As shown in Appendix \ref{app:rectsheet}, when we set $X=x-x'$, $Y=y-y'$ and $Z=z-z'$, the hyperkernel $\kc$ becomes 
\begin{flalign}
\nonumber
\kc &= -Z \atan \frac{XY}{Z |\vec{r}-\vec{r}'|}+Y \ln \frac{X+|\vec{r}-\vec{r}'|}{\sqrt{Y^2+Z^2}}\\
\nonumber
& \qquad +X \ln \frac{Y+|\vec{r}-\vec{r}'|}{\sqrt{X^2+Z^2}} \equiv \kc(X,Y,Z),
\end{flalign}
where $|\vec{r}-\vec{r}'|=\sqrt{X^2+Y^2+Z^2}$, and so the hyperpotential writes in the $3$D case
\begin{flalign}
\nonumber
\spsi(x,y,z) & = \iiint_\domain{\rho(x',y',z')} \\
& \qquad \times \kc(x-x',y-y',z-z')dx'dy'dz'.
\end{flalign}
Since $\rho=0$ outside $\domain$, the integral bounds can be safely changed for $\pm \infty$. We then conclude that {\it $\spsi$ is a convolution product}. This is expected because the potential itself is a convolution product \citep{binneytremaine87,haetal09}. We have
\begin{flalign}
\partial^2_{xy}( \rho \ast \kc) & = \rho \ast \partial^2_{xy} \kc \\
\nonumber
& = \rho \ast \frac{1}{|\vec{r}-\vec{r}'|}.
\end{flalign}
This result is {\it independent of the coordinate system}, namely:
\begin{equation}
\spsi = 
\begin{cases}
\Sigma \ast \ka, \quad  \text{in $2$D},\\\\
\rho \ast \ka, \quad \text{in $3$D}.
\end{cases}
\end{equation}

\subsection{Results in Cylindrical and Spherical coordinates}
\label{subsec:cyl}

In curved geometries, there are apparently fewer options. The reason is that a few formula for the potential of canonical surfaces are missing yet, and it is hard to find closed-form expressions for the hyperkernel by direct integration of Eq.(\ref{eq:k0}). One can probably use a series representation instead, but any truncation is expected to produce an approximate potential. Surfaces $\qt=const$ of particular interest are (see Fig. \ref{fig:ccs} and Tab. \ref{tab:kappas})
\begin{enumerate}
\item in \underline{cylindrical coordinates}:
\begin{enumerate}
\item a piece of a hollow cylinder (surface $R=const$);
\item a meridional sheet (surface $\theta=const$);
\item a polar sector (surface $z=0$);\\
\end{enumerate}
\item in \underline{spherical coordinates}:
\begin{enumerate}
\item a piece of spherical shell (surface $r=const$);
\item a meridional sector (surface $\theta=const$);
\item a piece of cone (surface $\phi=const$).
\end{enumerate}
\end{enumerate}

For cases 1a, 2a, and 2c (with $\phi' < \frac{\pi}{2}$), the potential is apparently not known in closed-form; this would be helpful. We have no hyperkernel to propose. This question remains open. Case 1b is accessible since the meridional sheet is nothing but a rectangular sheet in the plane $(x,z)$, rotated counter-clockwise by an angle $\theta'$. The formula for $\krz$ can then be deduced from the Cartesian case (see Appendix \ref{app:ms}); this is Eq.(\ref{eq:ms}). Case 1c can also be treated since the potential of a polar sector has been derived in \cite{hure12}. The formula is reproduced in the Appendix \ref{app:ps}, and the hyperkernel $\kps$ is given by Eq.(\ref{eq:k0atheta}). Finally, case 2b is feasible since the meridional sector is a polar sector. We can then use the result for $\kps$ established in cylindrical coordinates and apply convenient rotations to derive $\krphi$. A more direct calculus is presented in Appendix \ref{app:msect}.

\begin{figure}
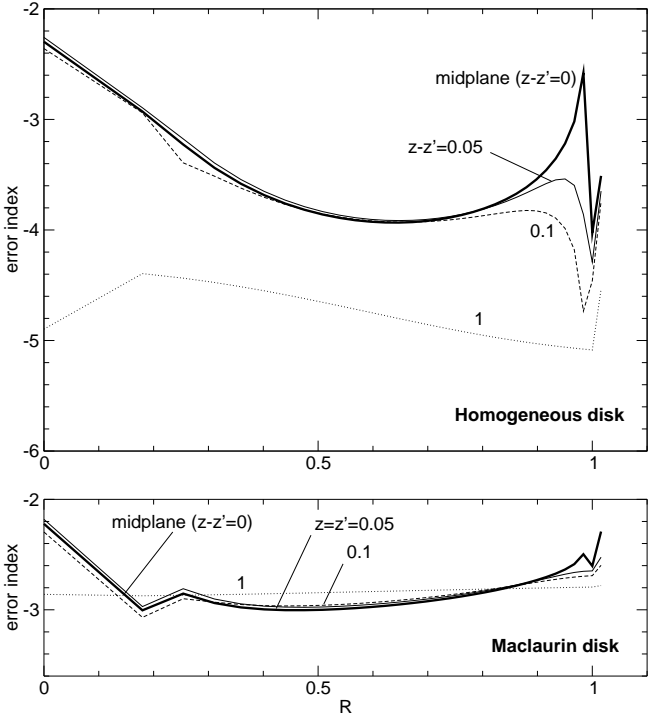

\centering
\includegraphics[width=8.5cm]{test_axidisc.eps}
\includegraphics[width=8.5cm]{test_aximcalaurindisc.eps}
\caption{Error index for the homogeneous disk ({\it top panel}), and for the (inhomogeneous) Maclaurin disk ({\it bottom panel}). The set-up is the same in both cases (edge at $R'=1$), and $N'=32$. The computational grid has $N=33$ points, but the same spacing.}
\label{fig:test_axidisc}
\end{figure}

\subsection{Axial symmetry}
\label{subsec:asym}

If the source is axially symmetrical (i.e., $\partial_{\theta'} \rho=0$), the integration of the Newton kernel in the P$'$-space over the polar angle $\theta' \in [0,2\pi]$ leads to
\begin{equation}
\int_{2\pi}{\frac{d\theta'}{|\vec{r}-\vec{r}'|}}= \frac{4}{\delta} \elik(k),
\label{eq:axi}
\end{equation}
where $\elik$ is the complete elliptic integral of the first kind, $\delta^2=(R'+R)^2+(z-z')^2$, and $k\delta=2\sqrt{RR'}$. Because this function is hyperbolically singular when $k \rightarrow 1$, we can determine an hyperpotential by an integration in the P-space. Again, there are not many options because we lack formulae for the potential of the hollow cylinder, for the cone, and for the piece of spherical shell (see Tab. \ref{tab:kappas}). Fortunately, there is the formula for the potential of the circular disk, i.e., a closed-form for $\int{\frac{4}{\delta} \elik(k) R'dR'}$ \citep{durand64, krough82, lassblitzer83,hure12}. We can therefore deduce an axially symmetrical hyperkernel $\kax$ by exchanging the role of P and P$'$ (see also Appendix \ref{app:ps}). This leads to Eq.(\ref{eq:k0_as}). The potential is also axially symmetrical, and it finally writes
\begin{equation}
\psi=\frac{1}{R}\partial_R \spsi,
\label{eq:psiaxi}
\end{equation}
where
\begin{equation}
\spsi = \int_\domain{\kax dm'}.
\label{eq:spsiaxi}
\end{equation}

We now present three last examples. Figure \ref{fig:test_axidisc} is for the circular disk with radius unity. The P-grid and the P$'$-grid are made of $N'=32$ and $N=N'+1$ points equally spaced in $R^2$ and $R'^2$ respectively | this radial scale is natural in this type of problem, both for the convolution and for the derivative. As above, the two grids coincide inside $\domain$ (there is just one point outside the disk) and the numerical schemes are second-order. The error index is shown at four different altitudes, including for the disk midplane (i.e. $z=z'$). The top panel shows the homogeneous disk. We see that the relative error is, on average, of about $0.1 \%$, with again a slight degradation near the edge. The bottom panel shows the Maclaurin disk where $\Sigma=\sqrt{1-R'^2}$. For this inhomogeneous case, the reference solution is taken from \cite{schulz09}. We see that the relative error is about $5 \times 10^{-3}$, which is not as good as in the homogeneous case. The error is almost insensitive to the altitude from the disk plane. This is due to the actual surface density profile (quadrature schemes are generally not very efficient to manage such situations).

\begin{figure}
\centering
\includegraphics[width=8.7cm,bb=70 50 320 264,clip==]{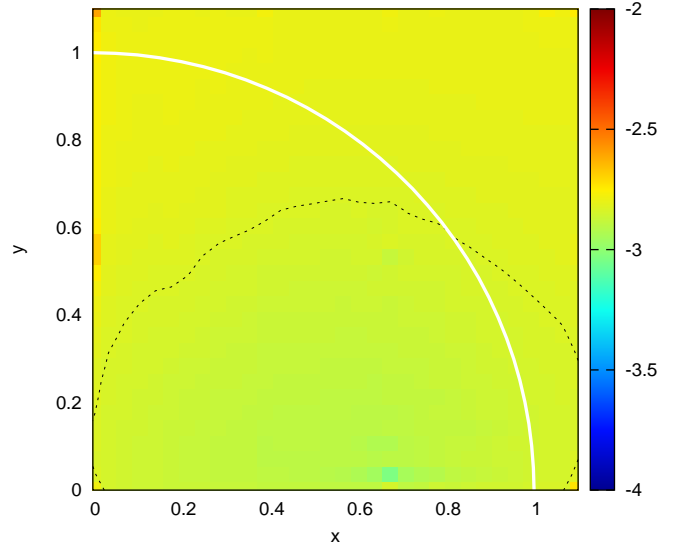}
\caption{Same legend and same color code as for Fig. \ref{fig:epsi_cart_cuboid}, but a radially inhomogeneous sphere (boundary in white; see text). The mean index is $-2.83$ (dashed line). The P-grid $(R,z)$ is a square, larger than the sphere's radius by $10\%$.}
\label{epsi_laneemden.ps}
\end{figure}

The last example is an inhomogeneous sphere with radius unity. The potential/density pair is the following
\begin{equation}
\label{eq:rhosph}
\begin{cases}
r' \le 1: & \rho(r')=\frac{\sin(\pi r') }{\pi r'},\\
&\psi_e(r) = -\frac{4G}{\pi}\left[1+\frac{\sin(\pi r) }{\pi r}\right]\\\\
r' \ge 1: & \rho(r')=0,\\
& \psi_e(r) = -\frac{4G}{\pi}\frac{1}{r}.
\end{cases}
\end{equation}
which corresponds to the solution of the Lane-Emden equation with polytropic exponent $\gamma=2$ (or index $n=1$), truncated at the first zero. The hyperpotential in spherical coordinates writes
\begin{equation}
\spsi = \int_0^\pi{ \sin \phi'  d\phi' \int_0^1{\rho(r')\kax r'^2dr'}}.
\end{equation}
where $\kax$ is the cylindrical hyperkernel $\kax$. This is a typical example where the surface $\pdomain$ (a disk) and the domain $\domain$ (concentric shells) are somewhat disconnected. Here, the sphere is discretized into $N' \times M'$ points equally spaced in the $(r',\phi')$-plane. Potential values are determined from Eq.(\ref{eq:psiaxi}) in the $(R^2,z)$-plane at $N \times M$ points equally spaced; the computational box is larger than the sphere's radius by $10\%$. Figure \ref{epsi_laneemden.ps} shows the error index for $N'=M'=32$ and $N=M=32$. We see that the deviation is remarkably homogeneous inside and outside the sphere. The relative error is about $0.2 \%$ inside and outside the material domain. This is comparable to the case of the cube.

Centrally symmetrical configurations can also be treated by using an hyperkernel, but there is nothing really new here (see Appendix \ref{app:centrals}).

\section{Summary and concluding remarks}

We demonstrated that the Newtonian potential of continuous bodies can be determined from the partial cross-differentiation of the mass density convolved with a finite amplitude kernel (a hyperkernel), regardless of any coordinate system. The recasting of Newton's integral is free of singularity and exact, and it applies to any type of two- and three-dimensional systems. Provided the hyperkernel is analytical, the extra-cost with respect to direct estimations is weak or negligible: it is only $N$ operations vs. $N^2$ in a grid with $N$ points. It is much lower if the method is only used to generate boundary conditions (and coupled with Poisson-solvers based on FFTs). The gain in accuracy is huge since i) direct estimates cannot avoid errors, ii) there is no free parameter, and iii) there is no trunctaed series. We have given a few examples in Cartesian, cylindrical, and spherical coordinates that prove the efficiency even with low-order quadrature and differentiation schemes. The recasting is therefore very attractive for any numerical applications. It is also a new tool for investigating various theoretical problems and derive potential/density pairs or approximations.

This work can therefore be continued and improved in several ways. Knowing analytical expressions for the hyperkernel associated with a given pair of orthogonal coordinates is the critical point of the method. We have shown that hyperkernels can be directly generated by considering the potential of homogeneous sheets, while there are doutblessly other possibilities. For Cartesian coordinates, all hyperkernels are known. In cylindrical and spherical geometries, a few closed-form expressions are apparently lacking yet (hollow cylinder and cone for instance). It would therefore be interesting to investigate this kind of question.  The formula for the polar sector should be helpful for most astrophysical applications however, such as for modelling rotating fluids. Other coordinate systems and geometries can be envisaged. For instance, the potential of inhomogeneous elliptic bodies can be determined, as done under central symmetry (see the end of Sect. \ref{subsec:asym}), from the theory of thin homeoids \citep[e.g.][]{macmillan1930theory}. This show the importance of seeking new potential/density pairs associated to $2$D-systems. It would also be interesting to analyze in more detail the numerical implementation of the method. There is obviously a wide panel of techniques at our disposal to perform quadratures/convolutions and differentiations, finite differences (as considered here), spectral methods, etc.

Finally, applications exceed the astrophysical context of gravitation. The approach is obviously suited to electrostatics and to electromagnetism since the potential vector is
\begin{equation}
\vec{A}(\vec{r}) = \int{\frac{\vec{u}dq'}{|\vec{r}-\vec{r}'|}},
\end{equation}
where $\vec{u}$ is the velocity of electric charges. It is also transposable to incompressible hydrodynamics where the pressure $p$ obeys a Poisson equation too,
\begin{equation}
p(\vec{r}) = - \int{\frac{ \nabla \cdot \left[ \left( \vec{u} \cdot \vec{\nabla} \right)  \vec{u} \right] dm'}{|\vec{r}-\vec{r}'|}},
\end{equation}
where $\vec{u}$ is the fluid velocity.

\begin{acknowledgements}
It is a pleasure to thank A. Dieckmann, M. Gazeau, F. Hersant, D. Pfenniger, and A. Pierens. I sincerely thank the referee for valuable scientific comments and advice to improve the organization of the paper.
\end{acknowledgements}

\bibliographystyle{aa}

\onecolumn
\appendix

\section{Hyperkernel for the rectangular sheet (Cartesian coordinates)}
\label{app:rectsheet}

The notations are those of Fig. \ref{fig:ccs}. Up to a factor $- \grav$, the potential of a homogeneous rectangular sheet with unity surface density is found from the integral
\begin{equation}
\iint{\frac{dx'dy'}{\sqrt{(x'-x)^2+(y'-y)^2+(z'-z)^2}}},
\end{equation}
where the bounds represents the coordinates of the four corners of the rectangular sheet. A closed-form expression is found for instance in \cite{durand64}.  If we set $X=x-x'$, $Y=y-y'$ and $Z=Z-z'$, we have $|\vec{r}-\vec{r}'|=\sqrt{X^2+Y^2+Z^2}$ and $dx'dy'=dXdY$. The indefinite integral is
\begin{flalign}
\iint{\frac{dXdY}{|\vec{r}-\vec{r}'|}}&= X -Y \ln \left(|\vec{r}-\vec{r}'|-X\right) -X \ln \left(|\vec{r}-\vec{r}'|-Y\right) -Z \left[ \atan \frac{X}{Z} + \atan \frac{XY}{Z |\vec{r}-\vec{r}'|} \right].
\end{flalign}
For the present prupose, we only need to generate the Newton kernel by a mixed partial derivative $\partial^2/\partial x \partial y$, and so, we have a certain liberty in choosing the most convenient integral bounds. Here, we take $x$ and $x'$ for the integral over $x'$, and $y$ and $y'$ for the integral over $y$. from this, we obtain
\begin{flalign}
\int_0^{x'-x}{dX\int_0^{y'-y}{\frac{dY}{|\vec{r}-\vec{r}'|}}}&=  -Y \ln \frac{|\vec{r}-\vec{r}'|-X}{\sqrt{Y^2+Z^2}} -X \ln \frac{|\vec{r}-\vec{r}'|-Y}{\sqrt{X^2+Z^2}} -Z \atan \frac{XY}{Z |\vec{r}-\vec{r}'|}.
\end{flalign}
Consequently, the hyperkernel is obtained from this expression by exchanging the variables $x'$ and $x$, and $y'$ and $y$. We find
\begin{flalign}
\label{eq:k0xy}
\kc&=  Y  \ln \frac{|\vec{r}-\vec{r}'|+X}{\sqrt{Y^2+Z^2}} + X  \ln \frac{|\vec{r}-\vec{r}'|+Y}{\sqrt{X^2+Z^2}}-Z \atan \frac{XY}{Z |\vec{r}-\vec{r}'|}.
\end{flalign}

To generate $\kappa^{yz}$ associated with a potential expressed as $\partial^2 \spsi /\partial y \partial z $, we perform the permutations $(x,x') \leftrightarrow (z,z')$ in the formulae above. To generate $\kappa^{xz}$ associated with a potential expressed as $\partial^2 \spsi/\partial x \partial z$, we perform the permutations  $(y,y') \leftrightarrow (z,z')$.

\section{Hyperkernel for the meridional sheet (cylindrical coordinates)}
\label{app:ms}

The notations are those of Fig. \ref{fig:ccs}. First, we use Eq.(\ref{eq:k0xy}) and perform the permutations $(y,y') \leftrightarrow (z,z')$. We find
\begin{flalign}
\kappa^{xz}&= Z  \ln \frac{|\vec{r}-\vec{r}'|+X}{\sqrt{Y^2+Z^2}} + X \ln \frac{|\vec{r}-\vec{r}'|+Z}{\sqrt{X^2+Y^2}}-Y \atan \frac{XZ}{Y |\vec{r}-\vec{r}'|}.
\end{flalign}
Then, we apply a counter-clockwise rotation (i.e., positive trigonometric sense) around the $z$-axis by an angle $\theta'$. We obtain the meridional sheet. The hyperkernel then writes
\begin{flalign}
 \nonumber
\kappa^{RZ}&= (z-z')  \ln \frac{|\vec{r}-\vec{r}'|-(R'+R \cos 2 \beta)}{\sqrt{R^2 \sin^2 2\beta+(z-z')^2}}  -(R'+R \cos 2 \beta) \ln \frac{|\vec{r}-\vec{r}'|+z-z'}{\sqrt{R^2+R'^2-2'R'R\cos 2\beta}}\\
& \qquad +R \sin 2 \beta \atan \frac{(z-z')(R'+R \cos 2 \beta)}{R \sin 2 \beta |\vec{r}-\vec{r}'|},
\label{eq:ms}
\end{flalign}
and the potential is given by
\begin{equation}
\psi(\vec{r}) = - \grav \partial^2_{Rz}{ \int_\domain{\krz dm'}}. 
\end{equation}

\section{Hyperkernel for the polar sector (cylindrical coordinates)}
\label{app:ps}

The notations are those of Fig. \ref{fig:ccs}. Again up to a factor $- \grav$, the potential of a circular sector is found from the formula
\begin{equation}
\iint{\frac{R' dR' d\theta'}{|\vec{r}-\vec{r}'|}},
\label{eq:intps}
\end{equation}
where
\begin{equation}
|\vec{r}-\vec{r}'|^2 = (R'+R)^2+(z-z')^2-4R'R \cos^2 \left(\frac{\theta-\theta'}{2}\right).
\end{equation}
This double integral has been calculated in \cite{hure12}. The indefinite form is
\begin{flalign}
\nonumber
\iint{\frac{R' dR' d\theta'}{|\vec{r}-\vec{r}'|}} & =\dd E(\beta,k) + \frac{R'^2-R^2}{\dd}F(\beta,k)  + \frac{R'-R}{R'+R}\frac{\zeta^2}{\dd} \Pi(\beta,m^2,k) - R \sin 2 \beta \asinh \frac{R'+R \cos 2 \beta}{\sqrt{\zeta^2+R^2 \sin^2 2 \beta}}\\& \qquad - \zeta \atan  \frac{\zeta(R'+R\cos 2 \beta)}{R \sin 2\beta \; |\vec{r}-\vec{r}'|},
\label{eq:potpolarsector}
\end{flalign}
where $F(\phi,k)$, $E(\phi,k)$ and $\Pi(\phi,m^2,k)$ are the incomplete elliptic integral of the first, second, and third kinds, respectively, $\delta=(R'+R)^2+\zeta^2$, $\zeta=z-z'$, $k\delta=2\sqrt{R'R}$, $2\beta=\pi - (\theta-\theta')$. To generate $\kps$, it is sufficient to consider the following integral bounds: $0$ and $R'$ for the radial integration, and $\theta-\pi$ and $\theta'$ for the angular part. Next, the source point and the field point are exchanged (note that $\delta$, $k$, $\zeta^2$ and $m$ are not impacted). We finally obtain
\begin{flalign}
\label{eq:k0atheta}
\kps&=\dd E(\beta',k) + \frac{R^2-R'^2}{\dd}F(\beta',k)  + \frac{R-R'}{R+R'}\frac{\zeta^2}{\dd} \Pi(\beta',m^2,k)\\ \nonumber
& \qquad + R'\sin 2 \beta \left(  \asinh \frac{R+R' \cos 2 \beta}{\sqrt{\zeta^2+R'^2 \sin^2 2 \beta}} - \asinh \frac{R' \cos 2 \beta}{\sqrt{\zeta^2+R'^2 \sin^2 2 \beta}} \right)\\ \nonumber
&  \qquad  \qquad + \zeta  \left\{\atan \left[ \frac{\zeta(R+R'\cos 2 \beta)}{a \sin 2\beta \; |\vec{r}-\vec{r}'|} \right] - \atan \left( \frac{\zeta}{\sqrt{\zeta^2+R'^2}} \cotan 2\beta \right) \right\},
\end{flalign}
where $2\beta'=\pi - (\theta'-\theta)=2\pi -2\beta$. The potential is then given by
\begin{equation}
\psi(\vec{r}) = - \frac{1}{R}\grav \partial^2_{R\theta}{ \int_\domain{\kps dm'}}. 
\end{equation}

Under axial symmetry, we have

\begin{flalign}
\label{eq:k0_as}
\kax=2 \left[ -\pi|\zeta| \epsilon' + \dd \elie(k) + \frac{R^2-R'^2}{\dd} \elik(k) + \frac{\zeta^2}{\dd}  \frac{R-R'}{R+R'} \elipi(m^2,k) \right],
\end{flalign}
where $\elie$, $\elik$ and $ \elipi$ are the complete elliptic integrals of the first, second, and third kinds, respectively.

\section{Hyperkernel for the meridional sector (spherical coordinates)}
\label{app:msect}

The notations are those of Fig. \ref{fig:ccs}. Up to a factor $- \grav$, the potential of a meridional sector defined by $\theta'=const$, is found from the double integral
\begin{equation}
\iint{\frac{\rp d\rp d\phi'}{|\vec{r}-\vec{r}'|}}.
\label{eq:intms}
\end{equation}
with convenient bounds. We can calculate this expression directly from the formula for the polar sector in cylindrical coordinates (see Appendix \ref{app:ps}). First, we write the relative separation $|\vec{r}-\vec{r}'|$ in the following form
\begin{equation}
|\vec{r}-\vec{r}'|^2 = (\rp +r \nu)^2+ r^2(1-\nu^2) -4 \rp r \nu \sin^2 \tau,
\end{equation}
where
\begin{equation}
\begin{cases}
\nu = \sqrt{1-\sin^2 \phi' \sin^2 2 \beta}\\
2 \beta = \pi - (\theta'-\theta)\\
2\tau = \pi - (\phi_0 -\phi)\\
\tan \phi_0 = - \tan \phi' \cos 2\beta.\\
\end{cases}
\end{equation}
Then, we notice that the similarity between Eq.(\ref{eq:intps}) and Eq.(\ref{eq:intms}) is perfect if we make the following substitutions
\begin{equation}
\begin{cases}
R'  \leftrightarrow \rp\\
R \leftrightarrow r \nu \\
(z-z')^2 \leftrightarrow r^2(1-\nu^2)\\
\beta \leftrightarrow \tau,\\
\end{cases}
\end{equation}
and then
\begin{equation}
\begin{cases}
\delta^2=r^2+\rp^2+2r\rp \nu,\\
k^2=\frac{4r\rp \nu}{\delta^2},
m^2=\frac{4r\rp \nu}{(\rp+r\nu)^2}.
\end{cases}
\end{equation}
From Eq.(\ref{eq:potpolarsector}), we see that Eq.(\ref{eq:intms}) becomes
\begin{flalign}
\nonumber
\iint{\frac{\rp d\rp d\phi'}{|\vec{r}-\vec{r}'|}} & =\dd E(\tau,k) + \frac{\rp^2-(r\nu)^2}{\dd}F(\tau,k)  + \frac{\rp-r\nu}{\rp+r\nu}\frac{r^2(1-\nu^2)}{\dd} \Pi(\tau,m^2,k) - r\nu \sin 2 \tau \asinh \frac{\rp+r\nu \cos 2 \tau}{r\sqrt{1-\nu^2 \cos^2 2 \tau}}\\& \qquad - r\sqrt{1-\nu^2}\atan  \frac{\sqrt{1-\nu^2}(\rp+r\nu\cos 2 \tau)}{\nu \sin 2\tau \; |\vec{r}-\vec{r}'|},
\label{eq:potmersector}
\end{flalign}
We then derive $\krphi=\iint{\frac{r dr d\phi}{|\vec{r}-\vec{r}'|}}$ from Eq(\ref{eq:k0atheta}) by making the same substitutions, and the potential is given by
\begin{equation}
\psi(\vec{r}) = - \frac{1}{r}\grav \partial^2_{r\phi}{ \int_\domain{\krphi dm'}}. 
\end{equation}

\section{Central symmetry}
\label{app:centrals}
When $\partial_{r'}\rho =0$, we can derive an hyperkernel by considering the potential of a spherical shell with unity surface density. From the Gauss theorem, we easily find (still up to $-\grav$)
\begin{flalign}
\int_\pi{\sin \phi' d\phi'\int_{2\pi}{\frac{d\theta'}{|\vec{r}-\vec{r}'|}}}&=\frac{4\pi}{r'} H(r'-r) + \frac{4\pi}{r} H(r-r'),
\end{flalign}
where $H$ is the Heaviside function. The hyperkernel $\kappa^{\rm shell}$ is then obtained by exchanging $r$ and $r'$ in this expression. In this case, we have simply $\psi = \spsi$ with
\begin{flalign}
\spsi & = - \grav \int_{r'}{\rho(r')r'^2 \kappa^{\rm shell}dr'} \\
\nonumber
      & = -4 \pi \grav \int{\rho(r')r'H(r'-r)dr'} - \frac{4\pi \grav}{r} \int{\rho(r')r'^2 H(r-r')dr'},
\end{flalign}
where the integral bounds are the inner radius and outer radius of the sphere. This result is well known \cite[e.g.][]{binneytremaine87}.

\end{document}